# Large collaboration in observational astronomy: the Gemini Planet Imager exoplanet survey case

Franck Marchis*[a], Paul G. Kalas[a,b], Marshall D. Perrin[c], Quinn M. Konopacky[d], Dmitry Savransky[e], Bruce Macintosh[f], Christian Marois[g], James R. Graham[b], and the GPIES Team

[a]Carl Sagan Center, SETI Institute, 189 Bernardo Avenue, Mountain View, CA, 94043, USA; [b]Dept. of Astronomy at UC Berkeley, Berkeley CA 94720, USA; [c] Space Telescope Science Institute, 3700 San Martin Drive, Baltimore MD 21218 USA; [d]Center for Astrophysics and Space Sciences, University of California, San Diego, 9500 Gilman Drive, La Jolla, CA 92093, USA; [e]Sibley School of Mechanical and Aerospace Engineering, Cornell University, Ithaca, NY 14853, USA; [f]Kavli Institute for Particle Astrophysics and Cosmology, Stanford University, Stanford, CA 94305, USA; [g]National Research Council of Canada Herzberg, 5071 West Saanich Road, Victoria, BC V9E 2E7, Canada

## ABSTRACT

The Gemini Planet Imager (GPI) is a next-generation high-contrast imager built for the Gemini Observatory. The GPI exoplanet survey (GPIES) consortium is made up of 102 researchers from ~28 institutions in North and South America and Europe. In November 2014, we launched a search for young Jovian planets and debris disks. In this paper, we discuss how we have coordinated the work done by this large team to improve the technical and scientific productivity of the campaign, and describe lessons we have learned that could be useful for future instrumentation-based astronomical surveys. The success of GPIES lies mostly on its decentralized structure, clear definition of policies that are signed by each member, and the heavy use of modern tools for communicating, exchanging information, and processing data.

## 1. INTRODUCTION

### 1.1 Historic of the Project

Like all research fields, astronomy has been radically transformed over the past two decades. Today, the daily work of astronomers is quite different from that done by previous generations, and radically different in the way that it is perceived by the general public. This is due largely to the accumulation of a large amount of data, the introduction of significantly more expensive and complex instruments, and the use of modern technology to communicate, process. and extract valuable information from the data.

The project was initiated as a concept in 2001 by a small group of astronomers led by Bruce Macintosh and James Graham. Following the success of adaptive optics systems that were being used at that time by large telescopes (W.M. Keck, VLT, Gemini) to image solar system bodies[1], study stellar populations[2], and reveal the true nature of the core of our galaxy[3], this team, composed mostly of members of the National Science Foundation's Center for Adaptive Optics (CfAO), started thinking about building an instrument able to image young Jupiter-like exoplanets. After studying concepts for several observatories, the team was selected in 2004 to carry out a conceptual design study of a exoplanet imager for the Gemini Observatory.

After seven years of development work, including a science meeting in 2007 (Fig 1), a preliminary design review in May 2007, a critical design review in May 2008, a procurement and fabrication phase until 2011, and a year and a half of integration at the UCSC LAO (Laboratory for Adaptive Optics) in 2013, the instrument, named the Gemini Planet Imager[4] (or GPI), was shipped to Chile in August 2013. GPI's first light occurred in November 2013[5] and science operation started in 2014.

Before 2014, the GPI team was composed largely of instrument-focused astronomers at a small number of institutions, most of which were located in California and Canada. In 2011, after a call by the Gemini Observatory to conduct a large survey with the instrument, the GPI team grew to include more observers, theorists, and junior astronomers and was officially selected to conduct a 3-year survey. The new team, called the Gemini Planet Imager Exoplanet Survey Consortium, began the survey in November 2014 to image and characterize young Jupiter-like exoplanets from a carefully selected sample of ~600 stars using 890 hours of telescope time between 2014 and 2017.

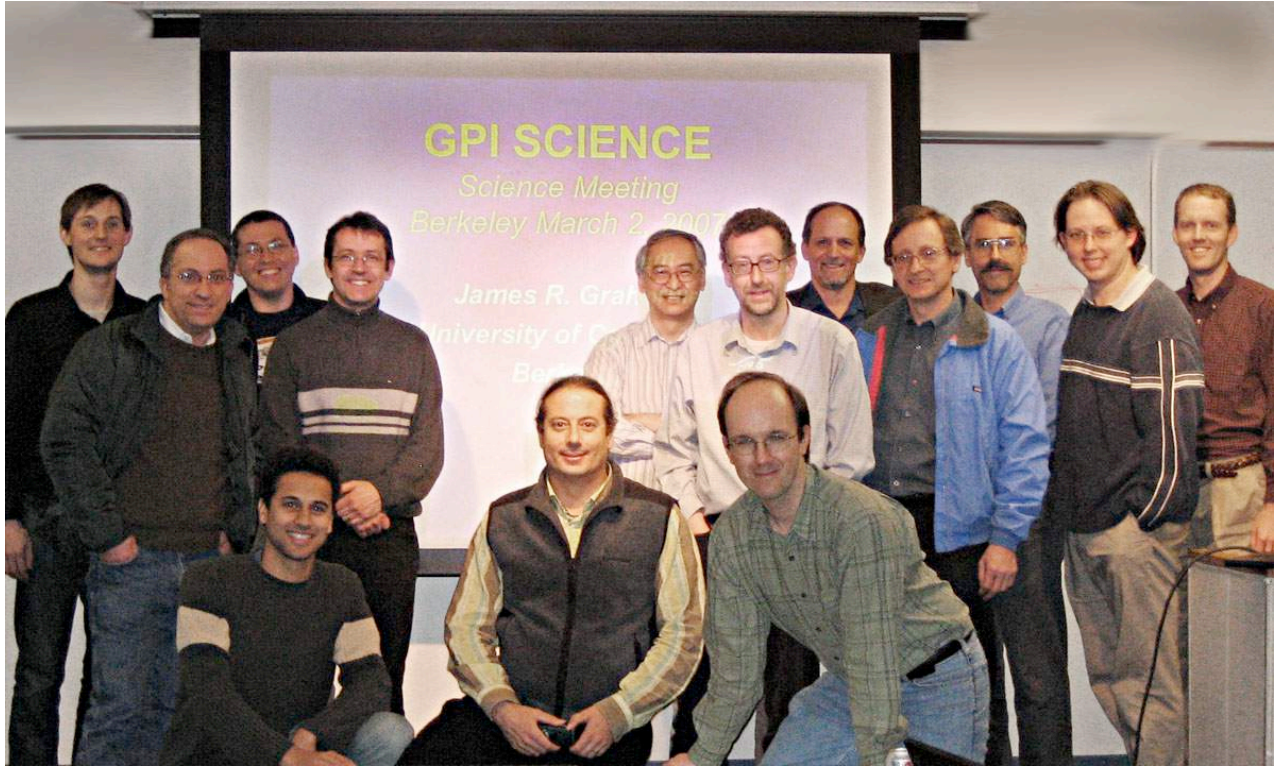

Figure 1. First Science Meeting of the GPI Instrument team in March 2007 at the University of California at Berkeley. Left to right: Marshall Perrin, Bijan Minati, Christian Marois, Franck Marchis, Rene Doyon, Paul Kalas, Michael Shao, James Graham, Bruce Macintosh, Geoff Marcy, Gene Serabyn, Dave Palmer, Paul Wright, Joe Jensen

### 1.2 Science Goals and Context

GPI is designed to directly detect the light from an extrasolar planet to determine its mass and composition, with an ultimate goal of determining the nature of our own planetary system. More than 3,400 exoplanets are known today but 79% were discovered by transit events that measure the planet's size and orbit, and by indirect Doppler technique (18%) that indicates the planet's mass and orbit. A few of them were directly imaged, a technique that provide an estimate of the planet's size, temperature, gravity, orbital parameters and even the composition of its atmosphere.

This search for exoplanets by direct imaging is a very competitive research area as illustrated by the similarities of two competing projects, SPHERE[6] and GPI, which:

- Are located in the same hemisphere (SPHERE is at the Mount Paranal in Chile and GPI is at Cerro Pachon also in Chile)

- Have roughly the same capabilities since both instruments are mounted on 8m-class telescopes (SPHERE is on UT4 of the Very Large Telescope, GPI is on the Gemini South telescope)

- Had their first light within 6 months (May 2014 for SPHERE and November 2013 for GPI) and started science operation very quickly after (October 2015 for SPHERE and November 2014 for GPI).

This collegial competitive environment forced the GPIES consortium to streamline its communication, data processing, data analysis and publication procedures despite the difficulties raised from a team spread across the northern American and Southern American continents and a diversity in seniority, personal interests and level of contribution to the project.

## 2. THE GPIES CONSORTIUM

### 2.1 Project Policies Document

In 2013, in anticipation of the complexity of the project and based on previous experiences of its members, several of us wrote a project-policies document that introduced consortium structure and rules. This detailed document includes a description of membership status, management structure, data and publication rights, rules on follow-up observations and collaborations, project teams and their roles, non-disclosure rules, and several scenarios on potential conflicts of interest. The consortium approved the document unanimously in November 2013 during our yearly science meeting.

At that time, we asked each consortium member to read and sign the document every year and agree to abide by its policies (Figure 2). We also reserved the right to modify the document, update sections, and add additional material to it as this became necessary.

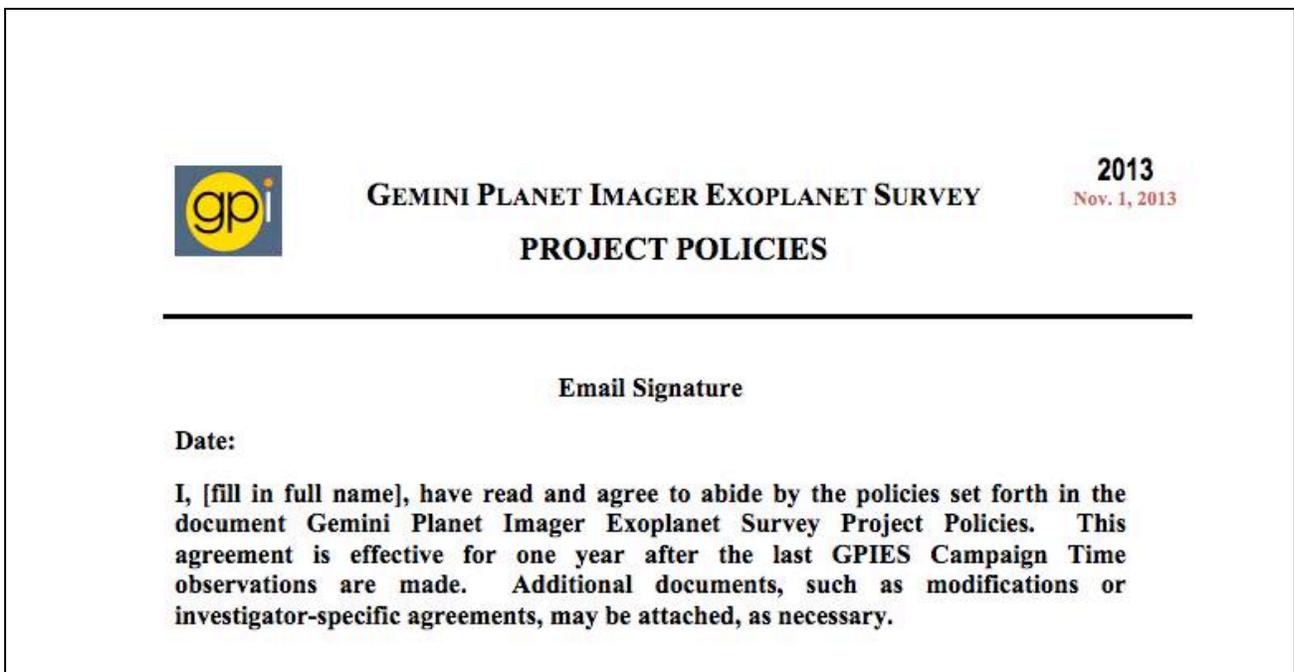

Figure 2. Last page of the Gemini Planet Imager Exoplanet Survey Project Policies document to be signed by each member.

### 2.2 Management

Management at GPIES is led by two bodies: the five-person **Executive Committee** (EC), which is responsible for administrative and policy matters, and the larger **Science Steering Committee** (SSC), which is responsible for the scientific direction of the project.

The project policies document states the following. The EC consists of five members selected to meet these criteria: time invested in the project, representativeness and expertise in the project, and work effort and/or funding brought to our work. The initial membership consisted of Bruce Macintosh and James Graham (permanent members), and Paul Kalas, Inseok Song, and Rene Doyon (rotating members with a two-year term). The EC nominates and elects its members by simple majority vote, rotating members as terms expire.

The EC is responsible for managing, coordinating, monitoring, and problem-solving the activities of the many scientists participating in GPI. The EC will make decisions by simple majority vote. The roles and responsibilities of the EC are:

(A) Coordinate the efforts of team leaders to insure that the campaign is carried out successfully, with a special emphasis on ensuring that all key deadlines are met.

(B) Raise and manage funding for science investigation. Funding is required to support all of our work, travel, hardware, and publications. PIs at different institutions control their individual funds—the EC and the SSC attempt to coordinate efforts such that duplication is avoided, all necessary tasks are funded, and proposals are do not conflict with one other. If a single coordinated proposal is made, e.g. to NSF, the EC will coordinate this effort.

(C) Resolve disputes. This includes appeals of any publication decisions made by the SSC, or if the SSC is unable to reach a consensus.

(D) Manage conflicts of interest, recusals, etc.

(E) Manage memberships and teams, such as approving new memberships or terminating existing ones. Both the SSC and EC must approve (by majority vote) any new collaboration members. The EC has sole authority to dismiss any collaboration members. Dismissal of the PI is made by a unanimous vote of both the EC and the SSC.

(F) As needed, select new team leaders. The EC is responsible for selecting team leads (and hence the SSC members) and has authority to replace a team lead if necessary, for reasons such as overcommitment or lack of responsiveness to the project's needs.

(G) Report to the Gemini Science Board and prepare any other project reports as needed.

(H) Coordinate efforts with competing external groups such as the SPHERE project.

(I) Participate in quarterly telecons to discuss the overall project status.

The Science Steering Committee (SSC) consists of the leaders of the separate project teams (described below) together with the members of the EC. The SSC is responsible for the overall scientific direction of the project, coordinating publication, etc. SSC decisions are made by simple or relative majority vote (What is a "relative majority vote"?). Members may appeal decisions made by the SSC to the EC. Such appeals must be made within two weeks of the decision, and the EC must respond within one week of the appeal. If no agreement is reached, the EC and the appealing member(s) will select a jointly acceptable three-person group of external mediators.

Specific SSC responsibilities include:

(a) Plan and manage science publications. Individual team leads will develop plans for publication, and the SSC will approve and coordinate them. Further details are found in Section III.

(b) Oversee, coordinate, and exercise ultimate authority over the dissemination of results, including conference presentations and press releases.

(c) Set the GPIES research agenda—the SSC is responsible approving the strategy of the campaign and any modifications to our research agenda in response to discoveries or a changing landscape

(d) Membership: both the SSC and the EC must separately approve new members by majority vote.

(e) Participate in monthly telecons and/or email discussions to review the scientific progress of the project.

## 2.3 Membership

In the years since the consortium was created and the GPI proposal was submitted, the number of team members has evolved continuously. In July 2016, the consortium was composed of one-hundred and two members from twenty eight institutions (Figure 3).

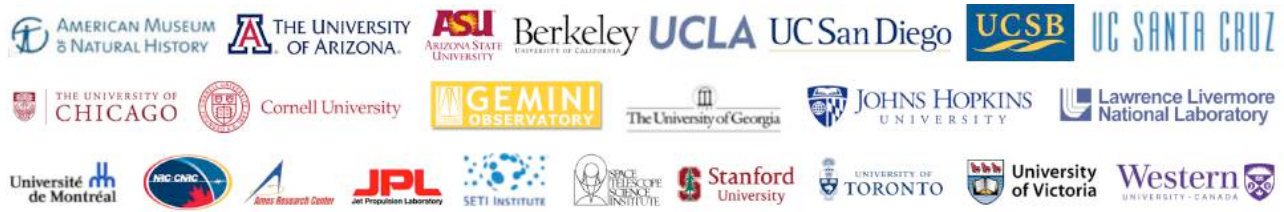

Figure 3. Logos of the main institutions, that hosts members of the GPIES consortium as of Aug 2015.

The project policies document sets the rules of membership. As stated in the document, membership has various levels that go from (A) to (E). Levels A-C receive greatest access to GPI scientific activities, and therefore must agree to non-disclosure of proprietary information.

(A) PI and Lead Co-I's: Primary members of the campaign science collaboration leading a team (See next section) within the project, generally identified at the time of proposal writing, and including the co-PIs from the instrument construction phase.

(B) CO-I's: Scientists with a significant role at the time of proposal writing and in the resulting collaboration.

(C) Programmatic Scientists: Typically postdoctoral researchers or graduate students, hired and annually sponsored by a level A or B members to work on GPIES and related tasks.

(D) Guest Scientists: Ad hoc members from outside the team brought in (sponsored) by a Level A or B member for specific expertise or a specific paper. Generally they are given access to only a subset of the campaign data. Guest scientists could also include people needed as co-I's, e.g., for access to another facility for follow-ups.

(E) Support Personnel: Programmers, technicians, engineers, etc.

Level A and B members retain their status even if they move to a new institution or position (but if they join a competing project, they may be required to sign a conflict of interest agreement or modify their GPIES membership level, including relinquishing it completely.) Level C members do not automatically remain in the project if they move to new institutions. An EC+SSC vote is required for them to remain, although the expectation is that in most cases this will occur as long as their new position allows them to continue to contribute to the project.

Promotion to level (B) membership requires a majority approval vote of the SSC and EC (Sometimes this is SCC+EC and sometimes it is "SCC and EC." I suggest picking one and using it uniformly.). Level C, D, and E memberships must be renewed annually in November by their Level A or B sponsor, who also has the authority to contact the SSC at any time to propose new members or terminate memberships that they sponsor.

Potential new members must be nominated and sponsored by an existing Level A/B (Same here—sometimes this is "A/B and other times it is "A and B." I would pick one and use it everywhere.) member. The nominee should compose a membership application that describes their qualifications, intended contributions, preferred roles, time commitment to the campaign, and any conflicts of interest. If the nominee's role closely fits with a specific team (see Section 2.4), the application should first be presented to the team leader(s) for review and possible modification. Otherwise the application should be sent first to the PI and project scientist for review and possible modifications. After the first review, the new membership application must be presented to the EC+SSC for discussion. After discussions with the SSC that may modify the details and scope of the Level C project, the EC+SSC will approve or reject the new Level C member and their proposed project by a simple majority vote.

Figure 4 shows the distribution in level of commitment of the GPI consortium.

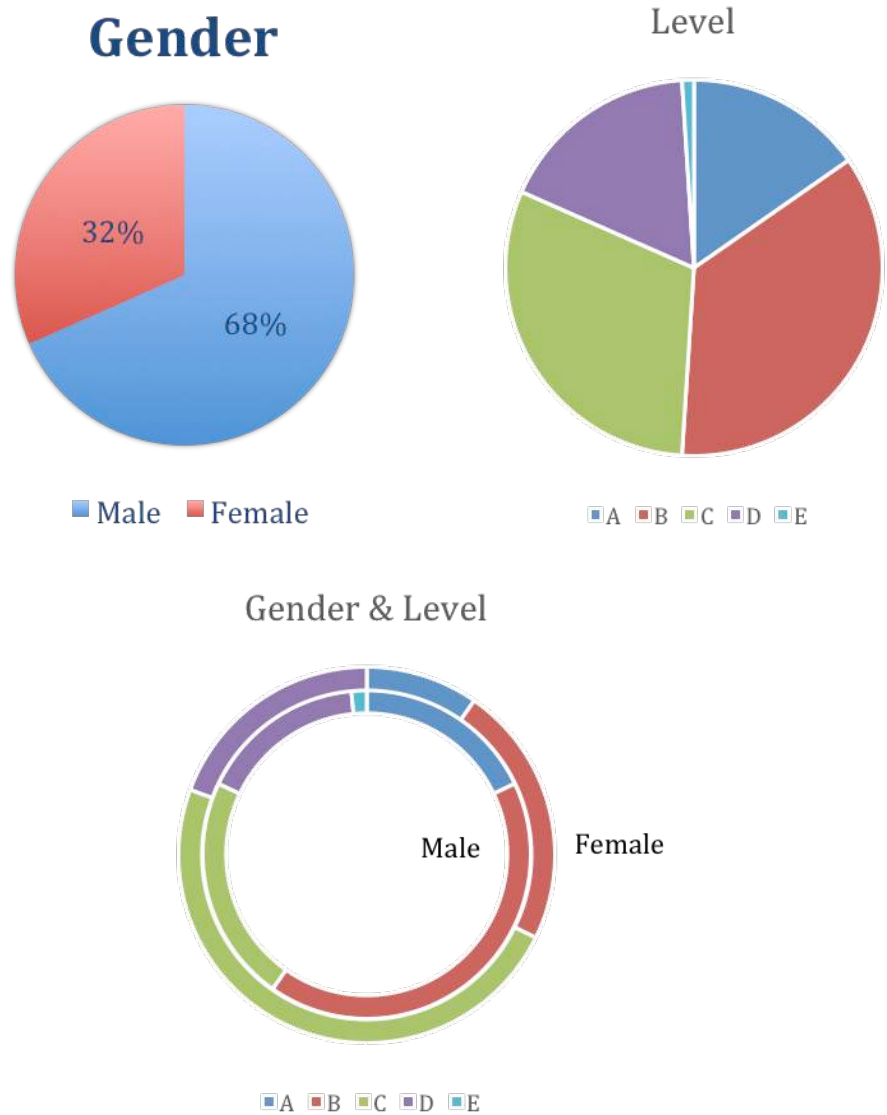

Figure 4. Diversity in gender and level of involvement of the GPI consortium as of July 2016.

**2.4 Project Teams and Roles**

We have divided the consortium into eight project teams (Figure 5) led by two members each. Consortium members can be on multiple teams. Each team has an email list to facilitate communication among members. Each team leader organizes communication within the team, including for example a weekly telecon, access to shared data and documents, and a wiki page. During the SSC telecon, the leaders of the team summarize their activity.

Here are our eight project teams and the tools and products generated by each:

- **Target Identification**:

    Role: Identify and characterize young exoplanet search targets; maintain a database of their properties.

    Tools and Products: Star catalogs from Simbad, previous surveys and unpublished observations.

- **Exoplanet Survey**:

    Role: Plan the exoplanet survey by selecting targets from the database to maximize scientific yield. Coordinate exoplanet science papers, particularly statistical analysis.

    Tools and Products: GPIES pipeline, analysis algorithms (pyKLIP,cADI, TLOCI, RDI, FMF), additional data.

- **Debris Disks:**

    Role: Identify debris-disk targets and ensure they are in the target database; coordinate debris-disk science, and conduct debris-disk modeling.

    Tools and Products: GPI pipeline, including polarimetry analysis mode, additional survey data.

- **Observing:**

    Role: Develop observing strategies (exposure times, filters, modes.) Coordinate prioritization of the target list for the Gemini queue. Prioritize and schedule follow-up observations. Primary point of contact with Gemini staff. Coordinate observing (if classical or remote) to insure coverage. Coordinate supplementary observations with other facilities.

    Tools and Products: Control remote room at Stanford U., nightly planning tool, access to processed (CADI, TLOCI,pyKLIP) data via SQL Query.

- **Data Analysis and Archiving:**

    Role: Maintain the survey data pipeline and data archives. Incorporate new features developed by other team members into the pipeline under appropriate software revision and version control. Coordinate extraction of quantitative data from images. Manage public data release at the end of the program.

    Tools and Products: Pipeline and manual, subversion code repository (Redmine), shared data (Dropbox), shared documents (Google Drive), Automated Data Processing System "Data Cruncher," Data analysis algorithms (pyKLIP, cADI, TLOCI, RDI), NERSC supercomputer

- **Exoplanet Modeling:**

    Roles: Provide modeling support for initial detection and determining properties of detected exoplanets from GPI data. Support papers on initial discoveries by providing relevant models and coordinate followup papers on modeling of individual objects.

    Tools and Products: utilizing spectra from GPI and spectral and photometric measurements from additional telescopes (W.M. Keck, Magellan) along with 1D radiative-convective equilibrium atmosphere models to produce model planet spectra for comparison with the observations. Also utilize retrieval methods to place constraints on planet properties.

- **Astrometry and Dynamics:**

    Roles: Coordinate astrometric calibration and extraction of data. Provide Keplerian orbit fits for detected objects. Model stability of multiplanet systems. Model evolution of debris disks..

    Tools and Products: Astrometric calibration of the instrument, GPI pipeline, MCMC and rejection sampling technique orbit fitting algorithms.

- **Communication, Education and Public Outreach:**

    Roles: Coordinate web sites, meeting tools, mailing lists, EPO programs. Coordinate PR, including public website, graphics for press releases, coordination of press releases, etc. Insure campaign has broad public impact.

    Tools and Products: Twelve mailing lists (Mailman), one wiki, one web site (Wordpress), three telecons per week on average (Zoom, Blue Jean), Slack, yearly Science meetings, hack day, artist-in-Residence program

Team activities changed over the first two years of the survey to meet our priorities. For instance, the Target Identification team had a significant role at the beginning of the survey, and then became dormant. Very recently, we were discussing mid-term survey strategy and reactivated it. Communications and EPO were two separate teams at the beginning of the project, but are now merged because of their similarity.

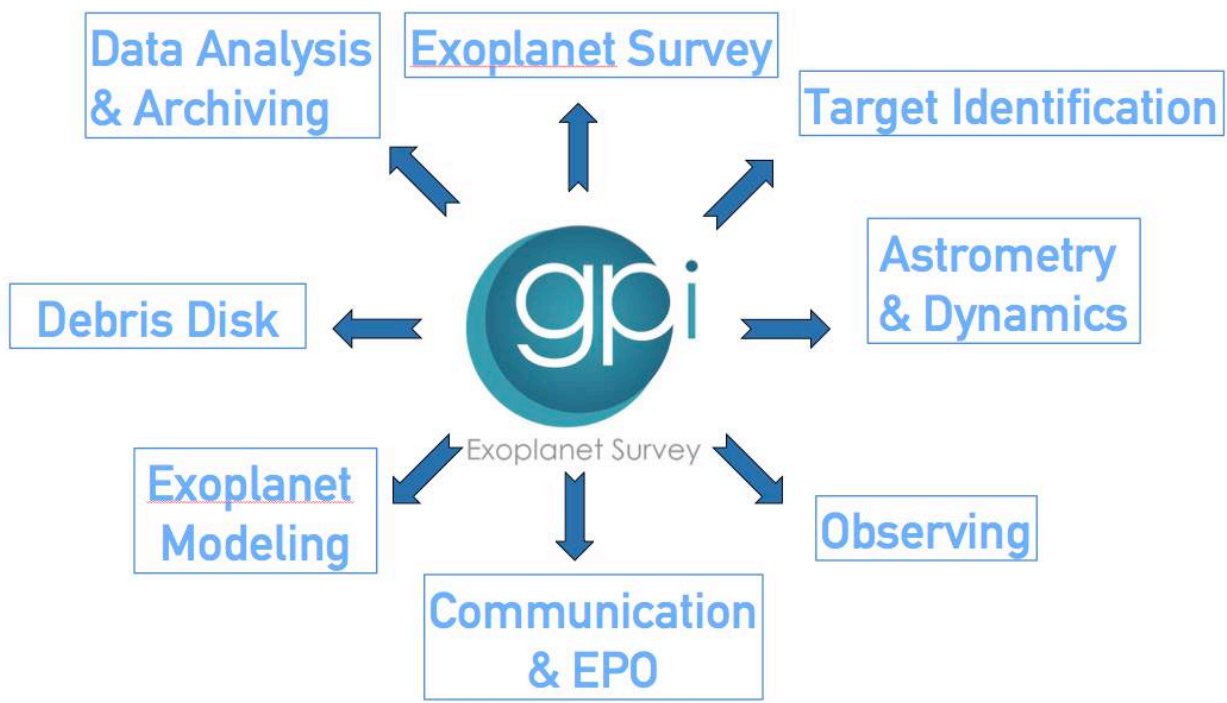

Figure 5. Project teams of the GPIES consortium.

## 3. LESSON LEARNED

### 3.1 GPIES in practice

GPIES productivity (See Section 4) has benefitted from a dedicated and effective team.
- "Dedicated" because for most GPI level A and B members are still working on their main scientific projects. Those team leaders have helped keep the project focused and on track by participating directly or working in their group to ensure success of the project, including development of the pipeline, rapid publication of key results, and communication of these results to the general public.
- "Effective" because before the beginning the survey we set rules for the consortium, spent a significant amount of time defining rules of publication (first authors, schedule), and kept our discussion transparent (e.g. our wiki includes a list of future publications and co-authors and can be used as a reference)

The Policies document evolved as the survey progressed:
- Several project teams changed, as have their team leaders
-  To clarify the rules of co-authorship, we established a builder list that consists of forty-seven members who have made significant contributions to construction of the original instrument, significant contributions to data analysis on multiple articles, construction of a key piece of campaign infrastructure, significant contribution to the target list, and been a member of the SSC and participated in four observing runs in Chile. All authors on the builder's list are included in alphabetical order at the end of the article's list of authors.
- As part of the ongoing recognition of these issues among the astronomical community, we are developing an anti-harassment policy in our policies document in September 2016 that we will discuss at our next science meeting to promote a safe and pleasant working environment for all members. These questions are particularly complex for multi-institution collaborations, where people from a wide variety of backgrounds and each answerable to a separate institutions may participate in an observing run.

It should be acknowledged that over the past ten years GPI consortium membership has diversified in age, gender, and the presence of traditionally under-represented populations in our work. Figure 6 is a recent (May 2015) group picture taken at our last GPI Science Meeting. The difference in diversity as compared to Figure 1 is obvious, and our current population better reflects the working force of the astronomy community. We do not have an official survey of the population of under-represented groups in the GPIES survey.

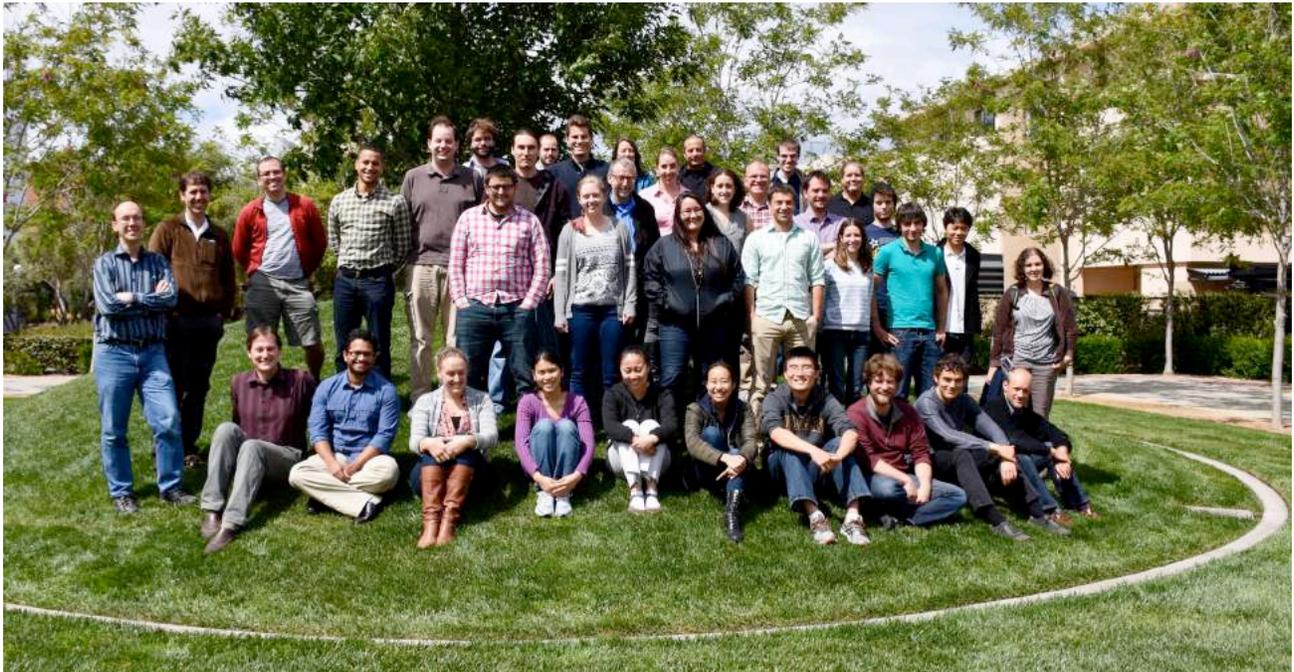

Figure 6. Group picture taken at our last GPI Science meeting at Stanford (May 2015).

**3.2  Is GPIES a truly decentralized scientific project?**

Traditional top-down organizations, including scientific ones, have adopted a highly centralized organization ("spider") that is not well fitted to our modern world and its rapid pace of change. Brafman and Beckstrom (2006)[7] have shown that decentralized organizations ("starfish") are better at surviving change because they adapt by restructuring themselves rapidly and effectively. When we designed the GPIES consortium management structure in 2011, we did not have a clear idea of the instrument's capabilities or its most successful science drivers, or much success in attracting funding and new members. Instead of setting the structural rules for the consortium, our PI and project scientist  have allowed the team to structure itself  and work together as issues and problems emerged.  Today, GPIES' structure is quite decentralized because it follows these four (of eight) major principles of decentralization:

- GPIES is an open system that does not have central intelligence; rather, intelligence is spread throughout the system. We do have a PI (Bruce Macintosh) and Project Scientist (James Graham); most of the key decisions are made after discussions with the SSC, or with the consortium itself
- GPIES is an open system that can easily mutate. We have done this over the years, transforming our structure as project teams became obsolete and adding more science cases to our investigations as appropriate.
- The GPIES consortium looks centralized but it is not.
- Put scientists into an open system like GPIES and they will automatically want to contribute.

Decentralized organizations share several characteristics such as the lack of a headquarters, a blurry division of roles, survivability if one of the teams or its leader is removed, a distribution of power and knowledge, a flexible management structure, a flexible number of participants, the existence of self-funded project groups, and finally the possibility for each participant to communicate directly with any other(s). All of these traits are seen in the GPIES consortium, which shows that indeed it could be one of the world's first "starfish" scientific organizations.

## 4. GPIES IMPACT IN OUR COMMUNITY

Even though the GPIES consortium was created relatively recently (2011) and started activity in 2014 with the kick-off of our campaign, it has had a significant impact on the astronomy community.
In July 2016, sixty articles and abstracts reporting GPIES results, including twenty-four peer-reviewed articles, have been published and more are coming.

Since 2011, three graduate students have received their PhDs based on GPIES data and eight members of the consortium were hired as university faculty or as staff in research institutes.

GPIES has also had a significant impact in the press, with more than 200 published articles on the instrument, its scientific results, and its team which have been published in news outlet around the world..

Finally, GPIES has created a team of skilled astronomers in high-contrast imaging who will use their expertise in future projects such as HABEX, LUVOIR or WFIRST. A significant part of the team that is designing the high-contrast imager and spectrograph for WFIRST is coming from the GPI consortium. Similarly from the ground, future proposed instruments such as MICHI and PSI for the TMT, as well as a balloon-based mission (MAPLE) concept, also include a significant part of the GPIES consortium. The direct imaging of exoplanets and the study of their atmosphere has become one of the main science drivers of ground-based and space-based telescopes. Projects like GPI and SPHERE have opened the path toward an even-more challenging problem, the detection of Earth-like exoplanets, which today seems accessible to us.

# Acknowledgements

The GPI project has been sup- ported by Gemini Observatory, which is operated by AURA, Inc., under a cooperative agreement with the NSF on behalf of the Gemini partnership: the NSF (USA), the National Research Council (Canada), CONICYT (Chile), the Australian Research Council (Australia), MCTI (Brazil) and MINCYT (Argentina). F.M. is supported in part by the NASA grant NNX14AJ80G